\documentstyle[12pt]{article}

\begin{document}
\pagestyle{empty}
\begin{flushright}
{BROWN-HET-961} \\
{to appear in MNRAS}
\end{flushright}
\vspace*{5mm}
\begin{center}
{\bf Statistics of Peculiar Velocities from Cosmic Strings} \\
[10mm]
R. Moessner\footnote{e-mail address moessner@het.brown.edu ; address from
October: Max Planck Institut f\"ur Astrophysik, Karl-Schwarzschild-Str. 1,
85740 Garching bei M\"unchen, Germany }\\
[10mm]
Department of Physics \\
Brown University \\
Providence RI. 02912. \\
USA. \\[2cm]

{\bf Abstract}
\end{center}
\vspace*{3mm}
We calculate the probability distribution of a single component of
peculiar velocities due to cosmic strings, smoothed over regions
with a radius of several $h^{-1}$ Mpc. The probability distribution is
shown to be Gaussian to good accuracy, in agreement with the
distribution of peculiar velocities deduced from the 1.9 Jy IRAS
redshift survey. Using the normalization of parameters of the cosmic string
model from CMB measurements, we show that the rms values for peculiar
velocities inferred from IRAS are consistent with the cosmic string model
provided that long strings have some small-scale structure.   \\

key words: cosmic strings ; large-scale structure of Universe ;
           cosmology:theory

\newpage
\setcounter{page}{1}
\pagestyle{plain}

\newcommand{\hri}{H^{-1}_{i}}
\newcommand{\Ri}{R_{i}}
\newcommand{\ui}{u_{i}}
\newcommand{\pv}{p(v_{x})}

\section{Introduction}
\smallskip
\par

Cosmic strings might be responsible for the formation of large-scale structure
which is observed in the Universe today. They are topological
defects which form in a phase transition in the early universe.
Topological defect models of structure formation are an alternative to
inflationary models. In inflationary models one assumes the existence of a
scalar field which drives a phase of very rapid expansion in the early
universe, and quantum fluctuations produced during this phase turn into
classical density perturbations which grow by gravitational instability
into present-day structures. In topological defect models the defects,
which are present at all times including today (and could possibly be
seen directly some day), act as seeds around which structures form.
It is important to study theories from both classes of models in order
to see which predictions can distinguish between different models, and which
predictions are too generic, arising in models based on widely different
assumptions.

Among the topological defect models, cosmic strings were the first to
be investigated in more detail (Brandenberger 1991), recently textures have
also been
subjected to closer scrutiny (Turok 1991). The cosmic string model looks
promising
so far, since normalizations of the parameters occuring in the model
obtained from observation agree with each other (Bennet, Bouchet \& Stebbins
1992, Perivolaropoulos 1993a).
These obervations include galaxy redshift surveys and measurements of
the temperature anisotropies in the cosmic microwave background radiation
(CMB). Moreover, cosmic strings naturally produce filaments and planar
structures in the matter distribution (Silk \& Vilenkin 1984, Vachaspati 1986,
Stebbins et al. 1987), in encouraging agreement with recent
galaxy redshift surveys. Another nice feature of the cosmic string model
is that it works well in the context of hot dark matter (Brandenberger 1987),
for which there are
candidates known to exist (the $\mu$ and $\tau$ neutrinos), although they may
be
massless. In contrast, none of the candidates for cold dark matter
required in inflationary models has been detected.

Topological defect models have not been as closely investigated as
inflationary models, and more work is necessary to quantify the
predictions of the cosmic string model. It is especially important to
find features accessible to observations in which cosmic strings differ
from the class of inflationary models, and statistics sensitive to
these features. Recently, the probability distribution of peculiar
velocities has been put forward as a statistic which might be capable of
discriminating
between cosmic strings and inflation (Kofman et al. 1994). Here we show that
according
to the calculations in a simple string toy model,
this statistic cannot be used to differentiate between cosmic strings and
inflation.

The IRAS 1.9 Jy redshift survey was analyzed (Strauss et al. 1990 \& 1992,
Yahil 1991) to obtain a uniform galaxy-density map. The peculiar velocities
were reconstructed from it via a self-consistent iterative scheme assuming
linear biasing between the density fluctuations of galaxies and mass (Nusser et
al. 1991). Using these results, Kofman et al. (1994) determined the probability
distribution of a single component of peculiar velocities of regions smoothed
over several $h^{-1}$ Mpc, with the result that it is consistent with the
underlying distribution $\pv$ being a normal distribution. Their result for the
Gaussianity of the velocity distribution is still tentative because of the
limited volume
sampled, and because velocities were not measured directly but deduced from a
redshift survey.

Inflationary models predict a normal distribution for $\pv$, whereas some
deviation from gaussianity
is expected in the cosmic string model since individual strings impart
coherent velocity perturbations over extended regions, as we will see in the
following section. The question is, however, just how large this deviation
from gaussianity is and whether it is big enough to be detected by current
observations. For a certain region receives velocity perturbations from
many strings, and by the central limit theorem many nongaussian
perturbations can add up to a gaussian signal.
Scherrer (1992) has shown that for seed models the velocity field can be
very nearly Gaussian even if the density field is nongaussian. In a model
where randomly distributed point masses, all of the same mass, accrete matter
gravitationally in a universe dominated by hot dark matter, he found that a
very low seed density of less than $10^{-2} {\rm{Mpc}}^{-3}$ is required in
order
for $\pv$ to be nongaussian.

The main aim of this paper is to calculate $\pv$ within the cosmic
string model in order to quantify its departure from gaussianity, and
to find out if the IRAS results are also in agreement with this stringy
probability distribution $\pv$.
Previously, 3 dimensional rms velocities have been obtained analytically within
the cosmic string model (Vachaspati 1992, Perivolaropoulos \& Vachaspati 1994),
and numerical simulations have been
performed (Vachaspati 1992, Hara, M\"{a}h\"{o}nen \& Miyoshi 1993) to obtain
the probability distribution of peculiar velocities, and
it was found that the distribution is quite gaussian.
Here we present the first analytical calculation of $\pv$ , within a model for
the string network which has been previously employed to make predictions
about temperature anisotropies in the CMB (Perivolaropoulos 1993a \& 1993b,
Moessner, Perivolaropoulos \& Brandenberger 1994) and the magnitude of
peculiar velocities (Vachaspati 1992, Perivolaropoulos \& Vachaspati 1994). The
dependence of the deviation from
gaussianity on one of the parameters of the model, namely the number
$\nu$ of strings per Hubble volume in the strings' scaling solution,
is shown explicitly.

Our conclusion is that on scales of several $h^{-1}$ Mpc, $\pv$ from
strings deviates only slightly from a normal distribution. On the smallest
scales and for the smallest number of strings per Hubble volume, the largest
deviation from
gaussianity is expected. Performing a $\chi^{2}$ -test of the stringy $\pv$
for the smallest scale of $6h^{-1}$ Mpc given in Kofman et al. (1994), with
imaginary data binned in the same way as in that paper, but drawn
from an underlying gaussian distribution (worst case),
yields the result that the stringy probability distribution is in
agreement with these data. This agreement holds for all values of $\nu$ , the
number of long strings per Hubble volume in the strings' scaling solution,
including $\nu = 1$.
This shows that the hope expressed in Kofman et al. (1994) that the
scenario for the formation of large scale structure, where widely
separated strings accrete matter in wakes behind them, can be ruled out
using the statistic $\pv$ is not realized.

In the next section we will describe the cosmic string model and the
mechanism for the production of velocity perturbations. In the third
section we will describe the analytic model for the string network and
calculate the moment generating function for a single component of the
peculiar velocities averaged over several $h^{-1}$ Mpc. Then we will obtain
the probability distribution $\pv$  from this moment generating function,
and compare it with observations.

\section{Cosmic Strings and Structure Formation}
Cosmic strings are linear topological defects formed at a phase transition
in the very early universe (Vilenkin 1985).
Those originating in the symmetry braking of a Grand Unified Theory possess an
enormous mass per unit length $\mu$  ($G \mu \approx 10^{-6}$ , where $G$ is
Newton's constant), and they
can be responsible for the structures observed today. Strings can have no ends,
so they are formed as
either infinitely long strings or closed loops. After formation the
network of strings quickly evolves towards a scaling solution where the
energy density in long strings remains a constant fraction of the total
background energy density. This is achieved by intercommutations and
self-intersections
of the strings leading to the production of small loops, which then
decay by emitting gravitational radiation. In this way, some of the energy
input into the string network coming from the stretching of the strings due to
the expansion of the universe is transferred to the background. Long strings
are straight over distances of the order of the horizon, so
that the scaling solution can be pictured as having a fixed number $\nu$
of long strings per Hubble volume at any given time.

Initially loops were thought to make the dominant contribution to
structure formation. At distances much larger than their radius they act
as point masses and accrete surrounding matter (Turok \& Brandenberger 1986,
Stebbins 1986, Sato 1986). Improved cosmic string evolution simulations (Bennet
\& Bouchet 1988, Albrecht \& Turok 1989, Allen \& Shellard 1990) showed that
more of the energy density is in long strings, and they are therefore more
important, accreting matter in the form of wakes
behind them as they move through space (Brandenberger, Perivolaropoulos \&
Stebbins 1990, Perivolaropoulos, Brandenberger \& Stebbins 1990).
Spacetime around a long straight cosmic string can be pictured as locally flat,
but with  a deficit angle of $8\pi G \mu$ (Vilenkin 1985). Therefore a string
moving relativistically with  velocity $v_{s}$ imparts velocity perturbations
to surrounding matter towards the
plane swept out by the string.
If small-scale structure is present on the string, there is in addition a
Newtonian force towards the string.
The magnitude of this velocity perturbation is
given by (Vachaspati \& Vilenkin 1991, Vollick 1992)
\begin{equation}
u=4 \pi G \mu \gamma_{s} v_{s} f ,
\;\;\;\;f=1+\frac{1-\rm{T}/\mu}{2(\gamma_{s}v_{s})^{2}}
\end{equation}
In the absence of small-scale structure on the string,the tension T of the
string is equal to its mass per unit length $\mu$, and $f=1$. If small-scale
structure is present, $\mu$ denotes the mass per unit length obtained after
averaging over the small scale structure, $\rm{T} \ne \mu$, and $f > 1$.
We consider the perturbations caused by strings after $t_{eq}$, the time of
equal matter and radiation, in a universe filled with hot or cold dark matter.
By the present time, the initial velocity perturbation imparted to the dark
matter at time $t_{i}$ has grown to (Brandenberger 1987, Stebbins 1987, Hara \&
Miyoshi 1990)
\begin{equation}
u_{i}\approx 0.4u\sqrt{z(t_{i})}
\end{equation}
Due to compensation (Traschen, Turok \& Brandenberger 1986, Veeraraghavan \&
Stebbins 1992), the deficit angle of strings which are straight over a horizon
distance, extends out only to a distance of one Hubble radius $H^{-1}$ from the
string, so that matter which is farther away does not receive any velocity
perturbations.
The velocity perturbation given in eq.(1) is independent of distance from
the string (up to the Hubble radius), so that cosmic strings impart
coherent perturbations over regions of the size of half a Hubble volume.

\section{Moment Generating Function}

The moment generating function (mgf) $M_{X}(t)$ of a random variable $X$ is
defined by
\begin{equation}
M_{X}(t)=\langle \exp^{tX} \rangle
\end{equation}
where the brackets denote the ensemble average, and it contains complete
information about $X$. In the following, $X = v_{x}$ denotes the random
variable for the component of the peculiar velocities  smoothed over a region
$\cal V$  of comoving radius $R$ in a fixed direction $\hat{e}_{x}$ .
We will calculate the mgf of $X$ in order to obtain the moments (and
cumulants) and probability distribution $\pv$ of $X$ from it.
The moments of $X$, $\mu_{j} = < X^{j} > $ are given by
\begin{equation}
\mu_{j}=\left(\frac{d^{j}}{dt^{j}} \right)_{t=0} M_{X}(t)
\end{equation}
and the cumulants $c_{j}$ are defined by
\begin{equation}
c_{j}=\left(\frac{d^{j}}{dt^{j}} \right)_{t=0}\ln(M_{X}(t))
\end{equation}
The probability distribution $p(v_{x})$ can be expanded in an asymptotic series
called Edgeworth series (Scherrer \& Bertschinger 1991, Stuart \& Ord 1987) in
terms of the quantities
\begin{equation}
\lambda_{j} = c_{j}/c_{2}^{j/2}
\end{equation}
and Hermite polynomials $H_{n}(x)$ defined by
\begin{equation}
H_{n}(x) = (-1)^{n} e^{x^{2}/2} \frac{d^{n}}{dx^{n}} e^{-x^{2}/2}
\end{equation}
For distributions with vanishing odd moments as in our case, the expansion is
\begin{equation}
p(\delta) = \frac{e^{-\delta^{2}/2}}{\sqrt{2} \pi} [1+ \frac{\lambda_{4}}{24}
H_{4}(\delta) +     \frac{\lambda_{6}}{720}H_{6}(\delta) +\frac{\lambda_{8} +
35 \lambda_{4}^{2}}{40320} H_{8}(\delta) + \cdots ]
\end{equation}
where $\delta =v_{x}/ \sigma$ , and $\sigma$ is the standard deviation of $X$.
Since it is an asymptotic expansion, the remainder is of the order of the last
term included (Erdelyi 1956).
The mgf has an important property which makes it useful for calculations.
For independent random variables $X$ and $Y$, the mgf of the sum is the
product of the individual moment generating functions
\begin{equation}
M_{X+Y}(t) = M_{X}(t) M_{Y}(t)
\end{equation}

The following calculations are carried out within an analytical model for the
string network
which has previously been used to obtain the temperature anisotropies in the
CMB and the magnitude of peculiar velocities from strings (see references given
in the introduction).
According to the scaling solution for cosmic strings, there is a fixed number
of long strings present per Hubble volume at any given time. After about one
expansion time of the universe (Hubble time), $t \rightarrow 2t$, these strings
will typically have self-intersected or intercommuted, so that the resulting
strings are uncorrelated with the ones at the previous Hubble time. We will
assume that during one expansion time $\nu$ long strings move across the Hubble
volume.
Each string is assumed to be straight over one horizon distance, and the effect
of all strings is taken to be the superposition of the effects of the
individual strings.
The fact that small-scale structure varies along strings is neglected, so that
we might somewhat underestimate the degree of non-Gaussianness of the velocity
distribution.
We also assume that the  strings' positions, velocities and orientations at
each Hubble time are random and uncorrelated, although this is not strictly
true, since - to mention one reason - the string network has the form of a
self-avoiding random walk.
According to the two previous assumptions, the random variable for the total
peculiar velocity, $X$, is the sum of independent random variables for the
velocity perturbations from the individual strings.
By eq.(9) we can therefore reduce the calculation of the the mgf for $X$ to
that of the mgf
for the contribution of only one string, and take the products afterwards.
In fact
we know that there are $\nu$ strings per Hubble volume on average. The products
in eq.(9) simplify if we take a Poisson distribution for the number of strings
per Hubble volume instead of assuming the presence of exactly $\nu$ of them
(Scherrer \& Bertschinger 1993). We can picture this as having a
reservoir of $n$ strings, each with a probability $\nu/n$ of being present in
a particular Hubble volume, in the limit that $n \rightarrow \infty$. So if
$M_{Y_{i}}(t)$ denotes the mgf for a single component of peculiar velocities
of the region $\cal V$ due to one string present at time $t_{i}$ in the
region's Hubble volume, then
\begin{eqnarray}
M_{X_{i}}(t) & = & \lim_{n \rightarrow \infty} [ \frac{\nu_{i}}{n} \langle
\exp^{tY_{i}} \rangle + ( 1-\frac{\nu_{i}}{n} ) ]^{n} \nonumber \\
             & = & \exp{[\nu_{i}(M_{Y_{i}}(t) - 1)]}
\end{eqnarray}
where $X_{i}$ is the random variable for the contribution of all strings at
Hubble time $t_{i}$ to the velocity perturbation of $\cal V$, and $\nu_{i}$
denotes the average number of strings having an effect on $\cal V$ at time
$t_{i}$, i.e. those strings which are within a distance of one Hubble radius
of $\cal V$ at time $t_{i}$ (see eq.(18)).

Since $X = \sum_{i=1}^{N} X_{i}$ ,
where $N$ is the number of expansion times since $t_{eq}$
\begin{equation}
N=\log_{2} \frac{t_{0}}{t_{eq}}
\end{equation}
and the $X_{i}$ are assumed to be independent,
\begin{eqnarray}
M_{X}(t) & = & \prod_{i=1}^{N} M_{X_{i}}(t)  \nonumber \\
         & = & \exp{[ \sum_{i=1}^{N} \nu_{i} ( M_{Y_{i}}(t) - 1)]}
\end{eqnarray}

We will now calculate the mgf for $Y_{i}$, the random variable for the
component
in the fixed direction ${\hat{e}}_{x}$ of the peculiar velocities of a region
$\cal{V}$ of comoving radius $R$ (the smoothing radius) due to one string
affecting the region at time $t_{i}$.  We have to perform the ensemble average
over positions , orientations ${\hat{e}}_{s}$ and directions of velocity
${\hat{v}}_{s}$ of the string.

For a long straight string only transverse velocities are observable, and we
assumed ${\hat{e}}_{s}$ and ${\hat{v}}_{s}$ to be random unit vectors.
Therefore
the unit normal $\hat{e}={\hat{e}}_{s} \times {\hat{v}}_{s}$ of the plane swept
out by the string, i.e. the direction in which matter receives velocity
perturbations, is itself a random unit vector. Consequently, the projection
$s = \hat{e} \cdot {\hat{e}}_{x}$ is uniformly distributed over the interval
$[-1,1]$, and the magnitude of the velocity perturbation from one string has to
be multiplied by $s$ to get the component in direction ${\hat{e}}_{x}$.

During one expansion time a string sweeps out a plane towards which matter
within a distance of one Hubble radius receives velocity perturbations, which
have grown to $u_{i}$ (see eq.(2)) by today. The possible values $y_{i}$ for
the random variable $Y_{i}$, the projection  of the peculiar velocity of
$\cal{V}$
in direction $\hat{e_{x}}$ due to one string at time $t_{i}$, are a function of
the perpendicular distance $r$ of the centre of $\cal{V}$ to this plane:
\[ y_{i}(r) = \;s \ui r/\Ri \;\;\;\; {\rm{for}} \;\;\; 0 < r < \Ri\]
\[ y_{i}(r) = \;s \ui \;\;\;\;  {\rm{for}}  \;\;\; \Ri < r < \hri - \Ri  \]
\begin{equation}
 y_{i}(r) = \;s \ui \frac{\hri-(r-\Ri)}{2 \Ri}  \;\; {\rm{for}} \;\;\hri-\Ri<
r<\hri+\Ri
\end{equation}
$\Ri$ is the physical size of the comoving radius $R$ at time $t_{i}$.
Strings within a distance of $H^{-1}(t_{i}) \equiv \hri$
of the region $\cal V$ can affect it. We distribute the centres $C$ of these
planes randomly within a sphere of radius $r_{max}^{i} = \hri$ around the
centre of $\cal{V}$. So the probability $p(c)$ for $C$ to be a distance $c$
from the
centre of $\cal{V}$ is
\begin{equation}
p(c) = \frac{3c^{2}}{(r_{max}^{i})^{3}}
\end{equation}
Since the normal of this plane has random direction, $r$ can be smaller or
equal to $c$, with probability
\begin{equation}
p(r;c) \approx 2 \frac{r}{c^{2}}
\end{equation}
Integrating over all $c$ gives the probability for the plane to be a distance r
from the centre of $\cal{V}$ as
\begin{equation}
p(r)= \int_{r}^{r_{max}^{i}} dc \; p(r;c) \; p(c) =
\frac{6r}{(r_{max}^{i})^{3}}
(r_{max}^{i} -r)
\end{equation}
The ensemble average thus becomes an integral over $r$ and $s$
\begin{eqnarray}
M_{Y_{i}}(t) & = & \langle \exp^{t Y_{i}} \rangle  \nonumber \\
             & = & \int_{0}^{r_{max}^{i}}  dr\;p(r) \int_{-1}^{1} ds \;
\frac{1}{2} \exp{(t y_{i}(r))}
\end{eqnarray}
These integrals can be done, and then eq.(12) can be used to obtain the mgf for
$X$ which includes the effect of all strings, with the number of strings
affecting $\cal V$ at time $t_{i}$ being given by
\begin{equation}
\nu_{i} = \nu(r_{max}^{i})^{3}/(\hri)^{3}
\end{equation}

There is one problem with the above. The formulas for $y_{i}(r)$ in equations
(13) are only true if the projection of $\cal{V}$ onto the plane swept out by
the string along its normal falls
completely into that plane, and is not (partly) outside of it. But the latter
can happen for large $c$ for some orientations of the plane, since one side of
the plane, in the direction of the string's motion, has a length  $l{_i} =
\hri$ or smaller, so that the distance of $C$ to the edge of the plane can be
smaller
than $\hri /2$. For $l{_i} = \hri$ one can show that less than half of the
strings miss $\cal{V}$ and give no perturbations to it, so that we can estimate
this effect by replacing $\nu$ by $\nu_{\rm{eff}} = 0.5 \nu$ .
If the strings are moving slowly, so that $l_{i}$ is even smaller, and not at
about the speed of
light, our model is not really applicable because the formula for the
imparted initial velocity perturbations would change.

Actually a string can affect $\cal{V}$ if $c \leq \Ri + \hri$ . But for $c$
larger than $\hri$ we encounter the problem mentioned in the previous
paragraph, so that we overestimate the perturbations by using $r_{max}^{i} =
\hri + \Ri$.
Therefore we calculate the cumulants for both $r_{max}^{i} = \hri$ and
$r_{max}^{i} = \hri + \Ri$ and take their average, and the model is more
accurate for smaller $R$. But since at scales below about $5h^{-1}$ Mpc
nonlinear effects become important, and we are only considering linear
perturbations, we must also keep above that scale.

\section{Probability Distribution and Comparison with Observations}

First we want to look at the shape of the probability distribution for $v_{x}$.
The nongaussianness
is largest on the smallest scales since smoothing makes things more
gaussian, and larger regions are affected by more strings. Therefore
we are going to compare $\pv$ from strings with the results from IRAS
at the smallest scale of $R = 6 h^{-1}$ Mpc considered in Kofman et al. (1994).
Also, the derivation of $M_{X}(t)$ is valid for scales of $R \leq l_{eq}$,
where $l_{eq} = 13 h^{-2}$ Mpc is the comoving size of the Hubble radius at the
time of equal matter and radiation.
The values $\Omega = 1$, $h = 1/2$ and $z_{eq} = 2.3 \cdot 10^{4} \Omega h^{2}$
are used.

Using a symbolic manipulation program (O'Dell 1991), the cumulants are
obtained from $M_{X}(t)$ according to eq.(5), giving the following values for
the $\lambda_{j}$ (eq.(6)) needed in the expansion of the probability
distribution (eq.(8)) for $R=6 h^{-1}$ Mpc :
\begin{eqnarray}
 \lambda_{4} & = & 0.34/\nu  \nonumber \\
 \lambda_{6} & = & 0.20/\nu^{2}  \nonumber \\
 \lambda_{8} & = & 0.15/\nu^{3}
\end{eqnarray}
These values, as well as the standard deviations quoted below, are the averages
of the two cases $r_{max}^{i}=\hri$ and $r_{max}^{i}=\hri + \Ri$ .
For two values of $\nu$, $p(v_{x}/\sigma)$  is plotted in Figure~1, up to
the term involving $H_{8}(\delta)$ in the expansion
(eq.(8)). For the higher value of $\nu = 10$ strings
per Hubble volume, the distribution is practically indistinguishable from
a gaussian one. For $\nu = 1$ there is a slight deviation. We want to
see if this deviation is significant by performing a $\chi^{2}$ -test with the
data given in Kofman et al. (1994), which has been grouped into bins of size
$v_{x}/\sigma = 0.25$. The data points fall practically on
a gaussian curve, so instead of taking the exact values from the data
we calculate the absolute frequencies $m_{j}$ in the $j$ bins expected if the
underlying
distribution were gaussian. The IRAS 1.9Jy survey maps out a sphere of radius
$80 h^{-1} \rm{Mpc}$ , so that there are $(80/6)^{3}$ independent smoothing
regions of radius $6h^{-1} \rm{Mpc}$. Let $n_{j}$ be the corresponding absolute
frequencies expected from the stringy distribution. Using 20 inner bins, we
find $\chi^{2} = 4.63$ for $\nu = 1$, where
\begin{equation}
\chi^{2} = \sum_{j=1}^{20} \frac{{(m_{j}-n_{j})}^2}{n_{j}}
\end{equation}
, which is much smaller than the $95 \%$ confidence upper limit of $38.58$ for
$19$ degrees of freedom, so that
the data is in agreement with the probability distribution from strings.
If we replace $\nu$ by $\nu_{\rm{eff}} = \nu / 2$ to take into account that
the side of the plane swept out by the string in the direction of its motion
is only half the diameter of the Hubble volume, then $\chi^{2} = 22.5$ for
$\nu = 1$.

Next we want to compare the magnitudes of velocities. The standard deviation
of the single velocity components is calculated to be
\begin{equation}
\sigma =  1.04 \nu^{1/2} \tilde{u} \;\;\; {\rm{for}}  \;\;R=6 h^{-1} \rm{Mpc}
\end{equation}
\begin{equation}
\sigma  =  0.99 \nu^{1/2} \tilde{u} \;\;\; {\rm{for}}  \;\;\;R=12 h^{-1}
\rm{Mpc}
\end{equation}
where $\tilde{u} = 0.4 z_{eq}^{1/2} u$ , and $u$ is defined in eq.(1).
We compare these values with results from Peacock and Dodds (1994),
who used the power spectra of various observations to calculate the
3 dimensional rms velocities $v_{rms}$ of regions of radius $R$. For a
gaussian random variable with three independent gaussian variables as
components, the standard deviation of a single component is given by
$\sigma = v_{rms}/ \sqrt{3}$. Using this relation, the values given in Peacock
and Dodds (1994) are
\begin{equation}
\sigma = (381 \pm 156) \; {\rm{km/s}} \;\;\; {\rm{for}}  \;\;R=6 h^{-1}
\rm{Mpc}
\end{equation}
\begin{equation}
\sigma  =  (337 \pm 138) \; {\rm{km/s}} \;\;\; {\rm{for}}  \;\;\;R=12 h^{-1}
\rm{Mpc}
\end{equation}
where we have taken the fractional error of $1/\sqrt{6}$ quoted for
the actual measurement of $v_{rms}$ at a scale of $5 h^{-1}$ Mpc.
Comparison of $\sigma$ in the string model with the values obtained from
observations at these two scales, gives as an average value for $\alpha f$
\begin{equation}
\overline{\alpha f} = 3.2 \pm 0.9
\end{equation}
where $\alpha$ is the combination of parameters
\begin{equation}
\alpha= \sqrt{\nu} \frac{G\mu}{10^{-6}} \frac{\gamma_{s} v_{s}}{c}
\end{equation}
$\alpha$ can be constrained  from the rms value of the temperature fluctuations
in the cosmic microwave background measured by COBE
to be (Perivolaropoulos 1993a)
\begin{equation}
\alpha = 1.0 \pm 0.2
\end{equation}
Using this value of $\alpha$ , we find
\begin{equation}
\bar{f} = 3.2 \pm 1.1
\end{equation}
This indicates that there must be some small scale structure on the strings
(see eq.(1)) in order to obtain the right magnitude of peculiar velocities and
consistency with CMB observations. This is also in agreement with recent
simulations (Bennett \& Bouchet 1988, Albrecht \& Turok 1989, Allen \& Shellard
1990), which show the presence of small-scale structure on cosmic strings. Our
analysis has been done in the string wake model, whereas strings with lots of
small scale structure accrete matter rather in the form of filaments. Therefore
the precise value of $f$ is not to be taken too seriously.

Numerical simulations of peculiar velocities from long strings without small
scale structure have been performed in a similar framework (Hara,
M\"{a}h\"{o}nen \& Miyoshi 1993), where
$\nu'$ strings are assumed to move across the horizon at every e-fold expansion
of the universe. The authors found that
\begin{equation}
\frac{G \mu}{10^{-6}} \frac{\gamma_{s}v_{s}}{c} \sqrt{\nu'} =( 4 \pm 1)
\end{equation}
yields good agreement with observations.
The number of strings at every two-fold expansion used in our analysis is
related to $\nu'$ by $\nu = \nu'\ln{2}$ . Therefore $\alpha f = 3.3 \pm 0.8 $
from these simulations,
which agrees quite well with our result of $\overline{\alpha f} = 3.2 \pm 0.9 $
{}.

\section{Discussion}
We have shown that the probability distribution of a single component of
peculiar velocities in the cosmic string wake model is very close to a normal
distribution on scales of several $h^{-1}$ Mpc, as suggested by a general
argument for seed models (Scherrer 1992), and in agreement with observations.
A comparison of the measured magnitude of peculiar velocities with that
expected from strings, using the normalization of string parameters from the
COBE quadrupole, suggested that strings have some small-scale structure.
Nongaussian features are more apparent in the velocity differences than in the
velocities themselves (Catelan \& Scherrer 1994), and it would be interesting
to calculate the
probability ditribution of these velocity differences within the cosmic string
model of structure formation.

\section{Acknowledgements}
I would like to thank Robert Brandenberger and Leandros Perivolaropoulos
for suggestions and helpful discussions. This work was supported in part by the
US Department of Energy under Grant DE-FG0291ER40688.

\newpage
{\bf References}

Albrecht A., Turok N., 1989, Phys. Rev. D 40 , 973.

Allen B., Shellard E.P.S, 1990, Phys. Rev. Lett. 64, 119.

Bennett D., Bouchet F., 1988, Phys. Rev. Lett. 60, 257.

Bennett D., Bouchet F., Stebbins A., 1992, ApJ (Lett.) 399, L5.

Brandenberger R., 1991, Phys. Scripta T36, 114.

Brandenberger R.,Perivolaropoulos L., Stebbins A., 1990, Int. J. Mod. Phys. A5,
1633.

Brandenberger R. et al., 1987, Phys. Rev. D36, 335.

Catelan P., Scherrer R., 1994, "Velocity Differences as a Probe of Non-Gaussian
Density \hspace*{.5in} Fields" (preprint SISSA Ref. 37/94/A).

Erdelyi A., 1956, "Asymptotic Expansions", (Dover).

Hara T., M\"{a}h\"{o}nen P., Miyoshi S., 1993, ApJ 415, 445.

Hara T., Miyoshi S., 1990, Prog. Theor. Phys. 81, 1187.

Kofman L. et al., 1994, ApJ 420, 44.

Moessner R., Perivolaropoulos L., Brandenberger R., 1994, ApJ 425, 365.

Nusser A. et al., 1991, ApJ 379, 6.

O'Dell J., 1991, "ALJABR", Fort Pond Research.

Peacock J., Dodds S., 1994, MNRAS 267, 1020 .

Perivolaropoulos L., 1993a, Phys. Lett. B 298, 305.

Perivolaropoulos L., 1993b, Phys. Rev. D 48, 1530.

Perivolaropoulos L., Brandenberger R., Stebbins A., 1990, Phys. Rev. D 41,
1764.

Perivolaropoulos L., Vachaspati T., 1994, ApJ 423, L77.

Sato H., 1986, Prog. Theor. Phys. 75, 1342.

Scherrer R., 1992, ApJ 390, 330.

Scherrer R., Bertschinger E., 1991, ApJ 381, 349.

Silk J., Vilenkin A., 1984, Phys. Rev. Lett. 53, 1700.

Stebbins A., 1986, ApJ (Lett.) 303, L21.

Stebbins A. et al., 1987, ApJ 322, 1.

Strauss M. et al., 1990, ApJ 361, 49.

Strauss M. et al., 1992, ApJ 385, 421.

Stuart A., Ord J., 1987, "Kendall's Advanced Theory of Statistics",
                 Vol.1 (London: \hspace*{.5in}Charles Griffin).

Traschen J., Turok N., Brandenberger R., 1986, Phys. Rev. D 34, 919.

Turok N.,1991, Phys. Scripta T36, 135.

Turok N., Brandenberger R., 1986, Phys. Rev. D 33, 2175.

Vachaspati T., 1986, Phys. Rev. Lett. 57, 1655.

Vachaspati T., 1992, Phys. Lett. B, 282, 305.

Vachaspati T., Vilenkin A., 1991, Phys. Rev. Lett. 67, 1057.

Veeraraghavan S., Stebbins A., 1992, ApJ Lett. 395, L55.

Vilenkin A., 1985, Phys. Rep. 121, 263.

Vollick D., 1992, Phys. Rev. D 45, 1884.

Yahil A. et al., 1991, ApJ 372, 380.

\newpage
\begin{center}
\bf Figure Captions
\end{center}
Figure 1: Stringy probability distribution $p(\delta= v_{x}/ \sigma)$ of a
single velocity component smoothed over regions of radius $R=6 h^{-1}$ Mpc  for
 for $\nu = 1$ (solid line) and
$\nu=10$ strings per Hubble volume (dotted line), compared with a normal
distribution (dashed line).

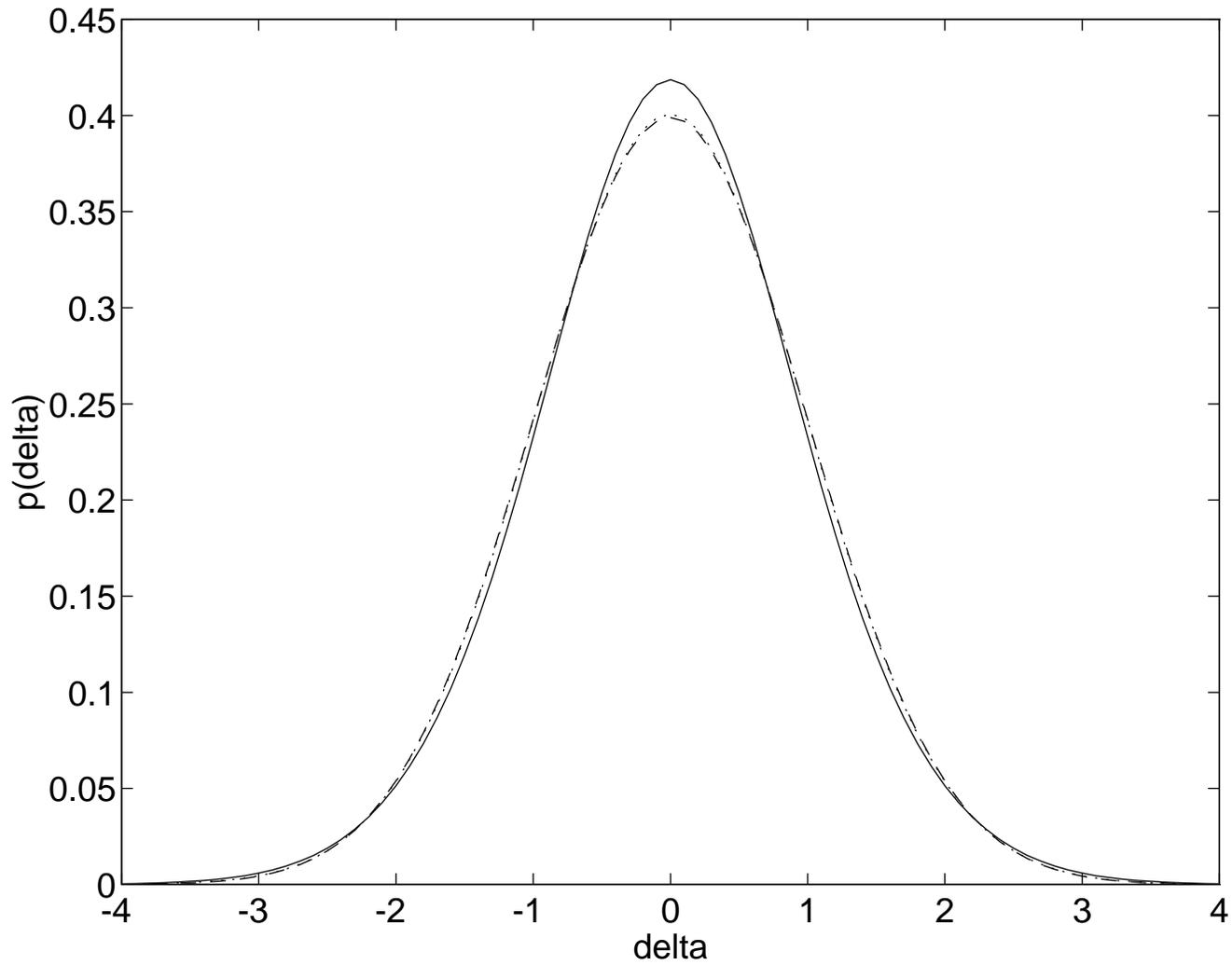
\end{document}